\documentclass{article}
\usepackage{spconf,amsmath,graphicx}
\usepackage{url} 

\title{Audio Mamba: Pretrained Audio State Space Model For Audio Tagging}
\twoauthors
 {Jiaju Lin}
	{Pennsylvania State University}
 {Haoxuan Hu}
	{Rutgers University}

\begin{document}
%
\maketitle
\begin{abstract}
Audio tagging is an important task of mapping audio samples to their corresponding categories. Recently endeavours that exploit transformer models in this field have achieved great success. However, the quadratic self-attention cost limits the scaling of audio transformer models and further constrains the development of more universal audio models. In this paper, we attempt to solve this problem by proposing Audio Mamba, a self-attention-free approach that captures long audio spectrogram dependency with state space models.
Our experimental results on two audio-tagging datasets demonstrate the parameter efficiency of Audio Mamba, it achieves comparable results to SOTA audio spectrogram transformers with one third parameters.

\end{abstract}
\section{Introduction}
\label{sec:intro}

Recently, with the success of pretrained-transformer\cite{vaswani2017attention, } models in language\cite{brown2020language, radford2019language} and vision learning\cite{dosovitskiy2020image,steiner2021train}, plenty of work explored introducing transformers in audio representation learning. With better performance and more unified architecture, the transformer-based model has replaced CNNs\cite{pann} and become the new trend in audio learning\cite{ast,htsat-ke2022}.
Despite the above advantages,  $O(n^2)$ cost of self-attention limits the application of transformers in long sequence modeling, such as audio spectrogram representation learning.
Some researchers have proposed some new $O(n)$ cost architectures to address this limitation, such as RWKV\cite{peng2023rwkv, duan2024vision} and space state models (SSMs) \cite{gu2021efficiently,fu2022hungry}. Among them, Mamba\cite{gu2023mamba，vmamba} is one of the newest attempts and demonstrates strong potential in both language and vision arenas. However, how to adapt Mamba into audio sphere is still under explored.
In this paper, we introduce the Mamba architecture for audio-tagging~\cite{bertin2011automatic}, which serves as the primary objective of pretraining endeavors. By exposing models to a diverse array of audio data and associating them with relevant tags or labels, audio pretraining equips these models with the ability to accurately recognize and categorize auditory stimuli, paving the way for more robust and effective audio processing systems. \par
In this work, we present the AudioMamba model, a neural network tailored for processing hierarchical audio representations. It employs Patch Embeddings~\cite{htsat-ke2022} and multi-stage SS Blocks~\cite{vmamba} with downsampling to refine feature extraction. Key components include depth-wise convolution~\cite{chollet2017xception}, SiLU activation~\cite{klambauer2017self}, and SS2D~\cite{vmamba} for adapting selective scanning to 2D data, enhancing both modularity and scalability. We utilize the AudioSet dataset, comprising over two million audio clips, to assess performance through metrics like mean Average Precision (mAP), mean Area Under Curve (mAUC), and d-prime. Training involves mono audio at a 32kHz sampling rate, transformed into mel-spectrograms. Our findings are substantiated by the open-source availability of the model’s implementation and weights, advancing audio processing research.\footnote{https://github.com/diggerdu/AudioMamba}.
Our contributions are: 
\begin{itemize}
    \item We propose AudioMamba, an architecture with linear time complexity to input sequence length. To our best knowledge, we are one of the first attempts that employ Mamba architecture for audio tagging. 
    \item We integrate the merits of HT-SAT and
    VMamba. AudioMamba is able to capture and process audio features at various scales by incorporating a multi-stage architecture and the specific patch embedding extraction approach.
    \item We conduct experiments on two audio-tagging datasets. The results illustrate the potential of SSMs in audio pertaining. AudioMamba achieves comparable performance with much fewer parameters. 
\end{itemize}

\section{Related Work}
\label{sec:relatedwork}

There are some studies directly related to our work. Ke Chen et al. proposed an audio transformer model designed for sound classification and detection tasks called HTS-AT\cite{htsat-ke2022}. Unlike existing audio transformers, HTS-AT\cite{htsat-ke2022} employs a hierarchical structure to reduce model size and training time while incorporating a token-semantic module for improved performance in event localization. The model demonstrates state-of-the-art results in various audio classification datasets and showcases better efficiency in terms of parameters and training duration compared to previous models.

Another work presented by Khaled Koutini et al. is Patchout\cite{patchout}, an innovative method for efficiently training and regularizing transformers on audio spectrograms, addressing the computational complexity challenges of transformers. The technique involves randomly dropping parts of the transformer input sequence during training. The proposed Patchout\cite{patchout} method, along with other complexity reduction strategies, achieves state-of-the-art performance on AudioSet, demonstrating improved performance and training efficiency compared to conventional CNNs and existing transformer models. The study highlights the potential of transformers in audio classification tasks, especially when computational resources are limited .

AST\cite{ast}(Audio Spectrogram Transformer) introduced by Yuan Gong et al., is a purely attention-based model for audio classification, challenging the necessity of convolutional layers in such tasks. The model leverages the Transformer architecture, adapted from vision to audio, to directly process spectrograms for classification. With the help of ImageNet pre-training, AST\cite{ast} achieves new benchmark results on datasets like AudioSet, ESC-50, and Speech Commands V2. The study underlines the potential of attention mechanisms, even in the absence of convolutional layers, for audio processing tasks.

Qiuqiang Kong et al. introduced Pretrained Audio Neural Networks (PANNs)\cite{pann}, which are trained on the large-scale AudioSet dataset and transferred to various audio pattern recognition tasks. Using the vast amount of training data available in AudioSet, PANNs\cite{pann} achieve state-of-the-art performance in several tasks, including audio tagging and acoustic scene classification. They demonstrated the effectiveness of transfer learning in audio processing and the significant impact of large-scale pretraining on the performance of audio neural networks. The Wavegram-Logmel-CNN architecture, which utilizes both waveform and log-mel spectrogram features, is highlighted for its superior performance in AudioSet tagging.

\section{Proposed AudioMamba Model}
\label{sec:model}
\begin{figure*}
    \centering
    \includegraphics[width=0.8\textwidth]{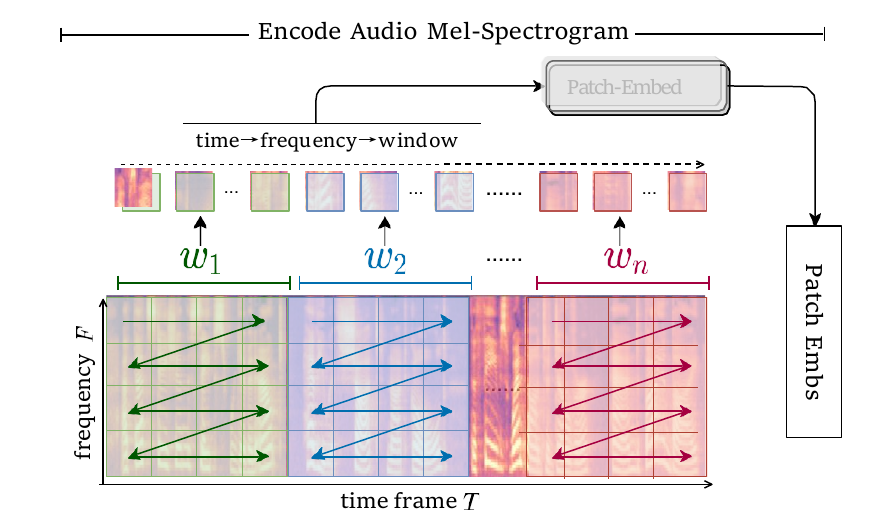}
    \caption{The patch embedding extraction process. The input spectrogram is first divided into a grid of non-overlapping patches. A Patch Embedding layer then maps each patch into a compact vector representation, capturing the local spectral features. These patch embeddings serve as the input to the subsequent SS blocks. This figure is a repurposed version of Figure 1 from our previous work~\cite{htsat-ke2022}.}
    \label{fig:input}
\end{figure*}

The proposed AudioMamba model is a novel approach to audio processing that leverages the power of hierarchical representations and state-of-the-art techniques in deep learning.  It's worth noting that while the original Mamba model is capable of handling audio-related tasks, it can only process waveforms directly and not the more commonly used audio spectrogram features. As an integration of HT-SAT~\cite{htsat-ke2022} and VMamba~\cite{vmamba}, AudioMamba adapts the data processing, training, and evaluation framework from HT-SAT while replacing the Swin Transformer used in HT-SAT with the state-space model employed in VMamba. By incorporating a multi-stage architecture and innovative components such as the SS Block, AudioMamba aims to capture and process audio features at various scales efficiently. This section delves into the key aspects of the AudioMamba model, starting with the encoding of the audio spectrogram and followed by an in-depth exploration of the model's backbone architecture. 

\subsection{Encode the Audio Spectrogram}
The described approach for extracting patch tokens from audio mel-spectrograms is a departure from the typical patch extraction method used for image-based vision transformers. In images, the patches are extracted in a grid-like fashion, as the width and height of the image both represent spatial dimensions.
However, in the case of audio mel-spectrograms, the width and height represent different information - the time and frequency bins, respectively. Additionally, the length of the time dimension is usually much longer than the frequency dimension. Therefore, simply applying a grid-based patch extraction may not effectively capture the relationship among the frequency bins within the same time frame.
To address this, the authors of~\cite{htsat-ke2022} propose a two-step patch extraction process . First, the mel-spectrogram is divided into \textbf{patch windows} $w_1, w_2, \dots, w_n$. Then, each window is further split into patches. The order of the resulting patch tokens follows \textbf{time $\rightarrow$ frequency $\rightarrow$ window}, ensuring that patches with the same time frame but different frequency bins are organized adjacently in the input sequence. This modification to the patch extraction process aims to better capture the relationships between frequency bins within the same time frame, which is an important consideration for audio-based sequence models.

\begin{figure*}
    \centering
    \includegraphics[width=1.0\textwidth]{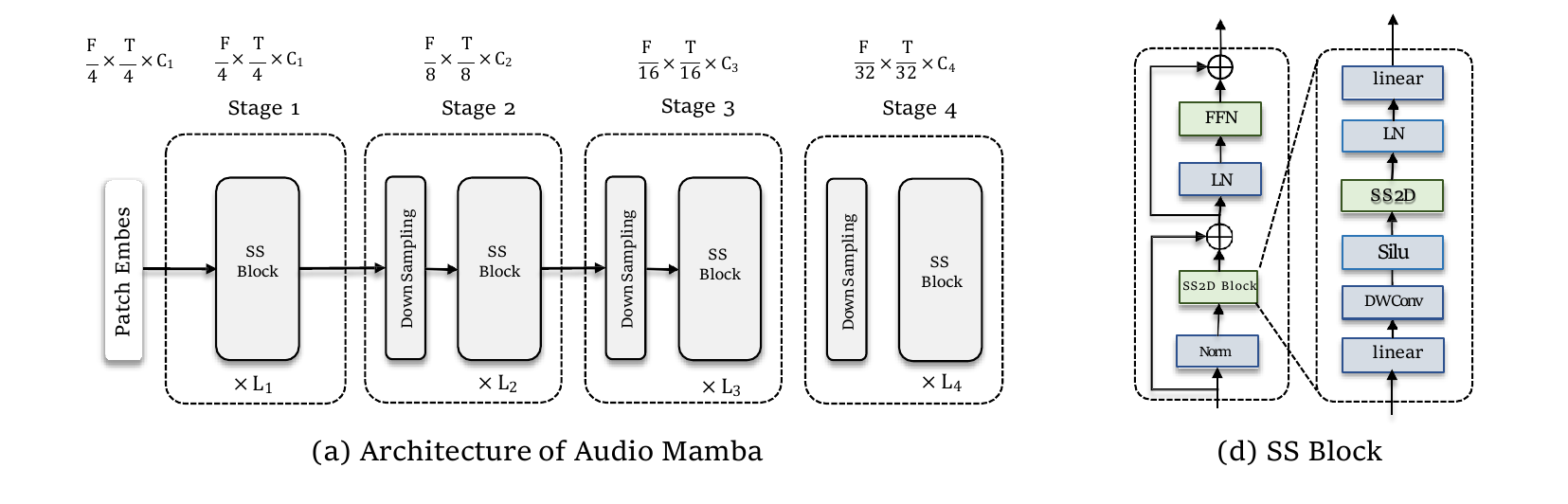}
    \caption{(a) Architecture of Audio Mamba, a multi-stage audio state-space model. The input patch embeddings undergo a series of stages, with each stage applying a block-based downsampling operation to progressively capture features at different scales. The outputs of downsampling layer are then processed through various neural network layers, including a Feature Fusion Network (FFN), Linear (LN) layers, and a Spatial Squeeze and Dimensional (SS2D) block, before producing the final output. (d) Details of the SS Block, which is a core component within the AudioMamba architecture. These patch embeddings serve as the input to the subsequent SS blocks. This figure is a repurposed version of Figure 8 from~\cite{vmamba}.}
    \label{fig:arch}
\end{figure*}

\subsection{Audio Mamba Backbone}

The Audio Mamba model, as shown in the provided figure~\ref{fig:arch}, is a neural network that aims to efficiently capture and process hierarchical audio representations through multiple stages. The model starts by converting the audio input into a suitable format for deep processing using Patch Embeddings. Within each stage, the model utilizes several SS Blocks to effectively extract and refine features. To progressively reduce the spatial dimensions while increasing channel depth, the model incorporates a downsampling mechanism via patch merging in all stages except the first, resulting in resolutions of $\frac{F}{8} \times \frac{T}{8}$, $\frac{F}{16} \times \frac{T}{16}$, and $\frac{F}{32} \times \frac{T}{32}$ at different stages.

The SS Block, a crucial component of the model, consists of a depth-wise convolution (DWConv)~\cite{chollet2017xception}, a Scaled Exponential Linear Unit (SiLU) activation~\cite{klambauer2017self}, and a unique operation called SS2D~\cite{vmamba}, which is essential for adapting the selective scanning technique to 2D data. The block's modular design incorporates linear transformations and normalization layers, ultimately leading to a Feed Forward Network (FFN) that enhances the model's feature extraction capabilities. This modular and scalable design of the SS Block enables the Audio Mamba model to efficiently handle complex audio processing tasks by improving adaptability and efficiency across the various stages of the network.

\section{Experiments And Results}
\label{sec:experiment}
\subsection{Training and Evaluation Details}
The AudioSet dataset more than two million audio clips, each lasting 10 seconds and categorized into 527 distinct sound event classes. Given that each audio clip contains two or more sound events, we formulate the modeling task as a set of 527 binary classifications.
 In this study, we adopt the training methodology described in HT-SAT~\cite{htsat-ke2022}, utilizing the entire training set, which includes 2 million samples, to develop our model and assess its performance on separate evaluation set consisting of 22,000 samples. We convert all audio samples to a mono format, employing a single channel with a sampling rate of 32kHz. For the generation of STFTs and mel-spectrograms, we apply a window size of 1024, a hop size of 320, and use 64 mel bins. Consequently, the dimensions of the mel-spectrogram are $(1024, 64)$, after padding each 10-second audio clip (equivalent to 1000 frames) with an additional 24 frames of silence, resulting in a total time frame length of $T=1024$ and frequency bins of $F=64$.

For the evaluation of different models on Audioset, we used mAP, mAUC and d-prime:
$$\text{mAP} = \frac{1}{Q}\sum_{q=1}^{Q}\sum_{k=1}^{n_q}\text{Precision}(R_{qk})\times\text{rel}(k)$$
mAP calculates the average precision across all classes, where $Q$ is the number of queries, $n_q$ is the number of retrieved items for query $q$, $\text{Precision}(R_{qk})$ is the precision at rank $k$ for query $q$, and $\text{rel}(k)$ is a binary indicator function that equals 1 if the item at rank $k$ is relevant, and 0 otherwise.
$$\text{mAUC} = \frac{1}{C}\sum_{c=1}^{C}\text{AUC}_c$$
mAUC is a metric used in binary classification tasks, where $C$ is the number of classes, and $\text{AUC}_c$ is the Area Under the Receiver Operating Characteristic (ROC) Curve for class $c$. 
$$d' = \frac{\mu_1 - \mu_2}{\sqrt{\frac{\sigma_1^2 + \sigma_2^2}{2}}}$$
$d'$ is calculated as the difference between the means of the signal ($\mu_1$) and noise ($\mu_2$) distributions, divided by the square root of the average variance of the two distributions ($\sigma_1^2$ and $\sigma_2^2$). \par


\subsection{Results Analysis}

\begin{table*}[htbp!]
\label{tab_audioset}
\centering
\renewcommand\arraystretch{1.1}

\setlength\tabcolsep{16pt}

\begin{tabular}{c|ccc|l} 
\hline
Model                             & \multicolumn{1}{c|}{mAP $\uparrow$} & \multicolumn{1}{c|}{mAUC $\uparrow$} & dPrime $\uparrow$       & Params $\downarrow$   \\ 
\hline
                                  & \multicolumn{3}{c|}{AudioSet}                                                                    &                    \\ 
\hline
HT-SAT (Cited)                    & 0.440                               & -                                  & -                     & 31M                \\
HT-SAT (Reproduced)               & 0.435                               & 0.950                               & 2.45                  & 31M                \\
PANN (Cited)                      & 0.434                               & -                                  & -                     & 81M                \\
DeepRes (Cited)                   & 0.392                               & -                                  & -                     & 26M                \\
Audio Mamba-Tiny                            & 0.440                               & 0.963                              & 2.51                  & 40M                \\
Audio Mamba-Micro & 0.437         & 0.962         &  2.46 & 12.3M              \\
Audio Mamba-Nano & 0.425         & 0.960         &  2.43 & 5.2M              \\

\hline
\end{tabular}

\caption{Performance of different models on AudioSet (- means that results are not shown in original papers.}
\end{table*}

The experimental section is divided into two parts. The first part involves pretraining the model on Audioset to assess its performance on this large-scale dataset. On average, each pretraining experiment requires 8-24 H100 GPU days. However, due to limited GPU resources, we conducted the second part of the experiments on the smaller ESC-50 dataset. This allows us to perform an ablation study to evaluate the effectiveness of various audio classification techniques when applied to the Audio Mamba model.

The table~\ref{tab_audioset} presents a detailed comparison of various audio classification models tested on the AudioSet dataset~\cite{audioset}, focusing on their performance metrics and model complexity.  It includes mean Average Precision (mAP), mean Area Under Curve (mAUC), and dPrime as performance indicators, with an additional column showing the number of parameters (Params) to reflect model size. Notably, both HT-SAT and PANN models are cited with their performance metrics, while Audio Mamba variants (Tiny, Micro, and Nano) are included to demonstrate scalability and efficiency variations in smaller models. Audio Mamba-Tiny achieves comparable performance to the best-cited model (HT-SAT) with slightly higher mAP and significantly improved mAUC and dPrime, albeit with more parameters. The Micro and Nano variants show a trade-off between reduced model size and slightly lower performance metrics, underscoring the effectiveness of the Audio Mamba architecture in maintaining competitive performance even with fewer parameters.

\begin{table*}[htbp!]
    \label{tab:ablation}
    \centering
    \caption{Ablation study on ESC50 dataset}
    \begin{tabular}{l|ccc}
        \hline
        Methods & \multicolumn{1}{c|}{f1-score(micro)} & \multicolumn{1}{c|}{f1-score(macro) } & Accuracy-score  \\
        \hline
        & \multicolumn{3}{c}{ESC50} \\
        \hline
        HT-SAT (from scratch) & 0.538 & 0.501 & 0.535 \\ 
                \hline
        Audio Mamba-Tiny & 0.435 & 0.408 & 0.435 \\
        $+$Transformer block & 0.438 & 0.407 & 0.438 \\
        $+$Imagenet Pre-train & 0.725 & 0.715 & 0.725 \\
        $+$Imagenet Pre-train $+$ cutmix & 0.728 & 0.716 & 0.728\\
        
        \hline
    \end{tabular}
    
\end{table*}
\subsection{Ablation Study}
In our pursuit to quantitatively evaluate the significance of various components within the Audio Mamba model, we designed a series of experiments centered on the ESC50 dataset~\cite{esc50}. The experimental settings, delineated in Table~\ref{tab:ablation}, encompassed diverse configurations aimed at illuminating the roles and impacts of specific elements.
 
\textbf{Baseline Establishment:} At the outset, we established a baseline by employing the Audio Mamba-Tiny model in its pristine form, without any alterations.

\textbf{Integration of Transformer Blocks:} Building upon insights gleaned from the Jamba framework~\cite{jamba}, which ingeniously merges Transformer and Mamba components to achieve a harmonious equilibrium among competing objectives such as memory efficiency, throughput maximization, and maintaining high quality, which introduced a notable enhancement. Specifically, we integrated a solitary transformer block post each SS block to gauge its efficacy and influence on model performance.

\textbf{Leveraging Pre-trained Models:} Drawing inspiration from recent advancements in audio-centric deep learning architectures, particularly AST~\cite{ast} and HST-AT~\cite{htsat-ke2022}, which have showcased substantial performance gains by leveraging pre-trained computer vision models, we embarked on a parallel exploration. To this end, we integrated a pre-trained VMamba model~\cite{vmamba} - tailored specifically for audio tasks - into our framework, thus probing the potential impact of pre-trained models on overall performance.

\textbf{Exploration of Data Augmentation Strategies: }Expanding our repertoire of investigation, we also delved into data augmentation strategies, recognizing their pivotal role in bolstering model robustness and generalization capabilities. Leveraging the pre-trained Vmamba model as a foundation, we explored the CutMix~\cite{cutmix} methodology. CutMix is an innovative augmentation strategy designed to enhance training efficacy and model performance in image classification tasks. CutMix involves the selective cutting and pasting of patches within training images, coupled with proportional mixing of ground truth labels. This approach optimizes the utilization of training pixels while retaining the regularization benefits of regional dropout.

Based on the findings presented in Table~\ref{tab:ablation}, it is observed that the inclusion of transformer blocks and the application of data augmentation techniques with cutmix method, exhibit minimal impact on the performance of Audio Mamba. However, the utilization of a pretrained Vmamba tiny model significantly enhances the results.

\textbf{Implications}: The HTS-AT model outperforms the Audio Mamba when trained from scratch on the ESC-50 dataset, yet on the AudioSet, Audio Mamba demonstrates superior performance compared to HTS-AT. This suggests that, compared to the Swin Transformer, the Mamba model may require more extensive data to achieve convergence in audio tasks. This could be attributed to the Swin Transformer's explicit modeling of local features, which may allow it to handle smaller datasets more effectively.

This section delineated our experimental methodology, encompassing baseline establishment, integration of transformer blocks, leveraging pre-trained models, and exploration of data augmentation strategies. Through meticulous experimentation, we aimed to elucidate the significance of various components within the Audio Mamba model, providing valuable insights for future optimization and advancement in Audio Mamba model.

\section{Conclusion}
\label{sec:conclusion}
In conclusion, the Audio Mamba model has demonstrated promising results in audio processing tasks, showcasing its capabilities on large datasets like AudioSet. As we continue to refine and expand upon this model, several avenues for future research have emerged. These are aimed at further enhancing the model's efficiency, applicability, and performance across various audio processing scenarios. Below are the key areas we plan to focus on in our future work. First, allocating additional GPU hours to explore the effectiveness of training tricks during the pretraining of AudioMamba on AudioSet. Besides, investigating the potential of AudioMamba for application in audio self-supervised training is another promising direction.
Also, how to fully exploit the potential of Mamba architecture is unexplored. For example, how to better utilize the intrinsic properties of audio, and what is the future universal architecture that achieves the best performance and applicability in diverse audio processing tasks.  

\bibliographystyle{IEEEbib}
\bibliography{strings}

\end{document}